# Bose-Einstein Condensation and Quasicrystals

M. Alexanian[1], V.E. Mkrtchian[2]

[1]*Department of Physics and Physical Oceanography,
University of North Carolina Wilmington, Wilmington, NC 28403-5606, USA*
[2]*Institute for Physical Research, Armenian Academy of Sciences, Ashtarak 0203, Republic of Armenia*

E-mail: alexanian@uncw.edu

**Abstract.** We consider interacting Bose particles in an external local potential. It is shown that large class of external quasicrystal potentials cannot sustain any type of Bose-Einstein condensates. Accordingly, at spatial dimensions $D \leq 2$ in such quasicrystal potentials a supersolid is not possible via Bose-Einstein condensates at finite temperatures. The latter also hold true for the two-dimensional Fibonacci tiling. However, supersolids do arise at $D \leq 2$ via Bose-Einstein condensates from infinitely long-range, nonlocal interparticle potentials.

## 1. Introduction

In a recent paper [1], the question of the existence of a Bose-Einstein condensate (BEC) in a supersolid was investigated. It was shown that an external crystalline lattice potential could not by itself sustain a condensate and so a crystalline lattice potential cannot give rise to a supersolid via a BEC. In addition, it was found that for spatial dimensions $D \leq 2$ self-crystallization occurs if the interparticle interaction between bosons is nonlocal and of infinitely long-range. In what following, we consider the same issues but now addressing quasicrystals, as well as, the 2-dimensional square Fibonacci tiling, which does not posses one of the "forbidden" $n$-fold rotational symmetries, $n \geq 5$, that are characteristic of quasicrystals and incompatible with translational periodicity.

## 2. Crystals

The Hamiltonian for the interacting Bose gas is

$$\hat{H} = \int d\mathbf{r}\hat{\psi}^\dagger(\mathbf{r})(\frac{-\hbar^2}{2m}\nabla^2)\hat{\psi}(\mathbf{r}) + \int d\mathbf{r}\hat{\psi}^\dagger(\mathbf{r})V_{ext}(\mathbf{r})\hat{\psi}(\mathbf{r})$$
$$+ \int d\mathbf{r}_1 \int d\mathbf{r}_2 \int d\mathbf{r}_3 \int d\mathbf{r}_4 \hat{\psi}^\dagger(\mathbf{r}_1)\hat{\psi}^\dagger(\mathbf{r}_2) V(\mathbf{r}_1,\mathbf{r}_2,\mathbf{r}_3,\mathbf{r}_4)\hat{\psi}(\mathbf{r}_4)\hat{\psi}(\mathbf{r}_3), \quad (1)$$

where $V_{ext}(\mathbf{r})$ is an arbitrary, external potential, $V(\mathbf{r}_1; \mathbf{r}_2; \mathbf{r}_3; \mathbf{r}_4)$ is a general two-particle interaction potential, and $\hat{\psi}(\mathbf{r})$ and $\hat{\psi}^\dagger(\mathbf{r})$ are bosonic field operators that destroy or create a particle at spatial position **r**, respectively.

Macroscopic occupation in the single-particle state $\psi(\mathbf{r})$ result in the non-vanishing [2] of the quasi-average $\psi(\mathbf{r}) = <\hat{\psi}(\mathbf{r})>$ and so the boson field operator

$$\hat{\psi}(\mathbf{r}) = \psi(\mathbf{r}) + \hat{\varphi}(\mathbf{r}), \quad (2)$$

where

$$\hat{\varphi}(\mathbf{r}) = \sqrt{\frac{1}{V(D)}} \sum_{\mathbf{k}} \hat{a}_{\mathbf{k}} e^{\mathbf{k} \cdot \mathbf{r}} \qquad (3)$$

with the condensate wavefunction

$$\psi(\mathbf{r}) = \sqrt{\frac{N_0}{V(D)}} \sum_{\mathbf{k}'} \xi_{\mathbf{k}'} e^{i\mathbf{k}' \cdot \mathbf{r}}, \qquad (4)$$

and normalization

$$\sum_{\mathbf{k}'} |\xi_{\mathbf{k}'}|^2 = 1, \qquad (5)$$

where $N_0$ is the number of atoms in the condensate, $V(D)$ is the D-dimensional "volume," $\hat{a}_{\mathbf{k}}^{\dagger}(\hat{a}_{\mathbf{k}})$ are the creation (annihilation) operators with commutation relations $[\hat{a}_{\mathbf{k}}, \hat{a}_{\mathbf{k}'}^{\dagger}] = \delta_{\mathbf{k},\mathbf{k}'}$, and $<\hat{\varphi}(\mathbf{r})>= 0$. The operator $\hat{\varphi}(\mathbf{r})$ has no Fourier components with momenta $\{\mathbf{k}'\}$ that are macroscopically occupied and so $\int d\mathbf{r}\hat{\varphi}^{\dagger}(\mathbf{r})\psi(\mathbf{r})$ = 0. The separation of $\hat{\psi}(\mathbf{r})$ into two parts gives rise to the following (gauge invariance) symmetry breaking term [2] associated with the interparticle potential in the Hamiltonian (1)

$$\hat{H}_{symm} = \int d\mathbf{r}_1 \hat{\varphi}^{\dagger}(\mathbf{r}_1) \int d\mathbf{r}_2 \int d\mathbf{r}_3 \int d\mathbf{r}_4 \psi^*(\mathbf{r}_2)[V(\mathbf{r}_1,\mathbf{r}_2,\mathbf{r}_3,\mathbf{r}_4) + V(\mathbf{r}_2,\mathbf{r}_1,\mathbf{r}_3,\mathbf{r}_4)]\psi(\mathbf{r}_3)\psi(\mathbf{r}_4) + h.c.$$
$$\equiv \int d\mathbf{r}_1 \hat{\varphi}^{\dagger}(\mathbf{r}_1)\chi(\mathbf{r}_1) + h.c. \qquad (6)$$

Recall that the interaction potential between bosons indicates that macroscopic occupation in a single-particle linear momentum state, viz., a spatially uniform condensate, does not give rise to additional macroscopic occupation in any other single-particle linear momentum states owing to the conservation of linear momentum by the interaction [1]. However, macroscopic occupation in two or more single-particle linear momentum states give rise to a denumerably infinite, macroscopically occupied states. For instance, macroscopic occupation in the single-particle momenta states **k**, **k**±**q**$_1$, and **k**±**q**$_2$, where **q**$_1$ x **q**$_2$ ≠ **0**, gives rise to additional macroscopic occupation in the single-particle momenta states **k** + $n_1$**q**$_1$ + $n_2$**q**$_2$, with $n_1$, $n_2$ = 0, ±1, ±2, • • • owing to the symmetry breaking term $\hat{H}_{symm}$.

Accordingly, the condensate wave function gets augmented and is of the Block form given by

$$\psi_{\mathbf{k}}(\mathbf{r}) = \sqrt{\frac{N_0}{V(D)}} \sum_{n_1,n_2=-\infty}^{\infty} \xi_{\mathbf{k}+n_1\mathbf{q}_1+n_2\mathbf{q}_2} e^{i(\mathbf{k}+n_1\mathbf{q}_1+n_2\mathbf{q}_2) \cdot \mathbf{r}}$$
$$\equiv e^{i\mathbf{k}\cdot\mathbf{r}} u_{\mathbf{k}}(\mathbf{r}), \qquad (7)$$

with $u_{\mathbf{k}}(\mathbf{r} + \mathbf{r}_0) = u_{\mathbf{k}}(\mathbf{r})$, where

$$\mathbf{r}_0 = 2\pi \left[ \frac{(q_2^2 - \mathbf{q}_1 \cdot \mathbf{q}_2)\mathbf{q}_1 + (q_1^2 - \mathbf{q}_1 \cdot \mathbf{q}_2)\mathbf{q}_2}{q_1^2 q_2^2 - (\mathbf{q}_1 \cdot \mathbf{q}_2)^2} \right]. \qquad (8)$$

### 3. Quasicrystals

We now consider the replacement (2) in the term in (1) associated with the external, local potential $V_{ext}(\mathbf{r})$. One obtains the symmetry breaking Hamiltonian

$$\hat{H}_{symm}^{(ext)} = \int d\mathbf{r} \hat{\varphi}^{\dagger}(\mathbf{r}) V_{ext}(\mathbf{r}) \psi(\mathbf{r}) + h.c. \qquad (9)$$

Consider the local, finite two-dimensional quasicrystal lattice potential,

$$V_{ext}(\mathbf{r}) = \frac{1}{(2\pi)^2} \sum_{\mathbf{k}} g(\mathbf{k}) \sum_{m_1 \cdots m_n = -M_1 \cdots -M_n}^{M_1 \cdots M_n} e^{-i\mathbf{k} \cdot \left(\mathbf{r} - \sum_{i=1}^n m_i(\alpha_i \mathbf{a} + \beta_i \mathbf{b})\right)} \tag{10}$$

where $g(\mathbf{k})$ is the Fourier transform, $\mathbf{a}$ and $\mathbf{b}$ are arbitrary two-dimensional vectors in the x-y plane with $\mathbf{a} \times \mathbf{b} \neq 0$, $(\alpha_i \mathbf{a} + \beta_i \mathbf{b}) \times (\alpha_j \mathbf{a} + \beta_j \mathbf{b}) \neq 0$, $\alpha_i$ and $\beta_i$ are irrational numbers, and $n \geq 3$. In (10), we have projected a periodic structure in $n$-dimensional space into a $D$-dimensional quasicrystal space ($n > D$). Cases $n = 1, 2$ reduce to a one- and two-dimensional crystals, respectively. One obtains that

$$\hat{H}_{symm}^{(ext)} = \int d\mathbf{r} \hat{\varphi}^\dagger(\mathbf{r}) V_{ext}(\mathbf{r}) \psi(\mathbf{r}) + h.c.$$
$$= \frac{\sqrt{N_0}}{V(D)} \sum_{\mathbf{k}_1, \mathbf{k}_2 \neq \mathbf{k}_1} \hat{a}_{\mathbf{k}_1}^\dagger \xi_{\mathbf{k}_2} g(\mathbf{k}) \prod_{i=1}^n \frac{\sin\left[\mathbf{k} \cdot (\alpha_i \mathbf{a} + \beta_i \mathbf{b})(M_i + 1/2)\right]}{\sin\left[\mathbf{k} \cdot (\alpha_i \mathbf{a} + \beta_i \mathbf{b})/2\right]} + h.c., \tag{11}$$

where $\mathbf{k} \equiv \mathbf{k}_2 - \mathbf{k}_1$, which follows with the aid of (3), (4), and (10). Recall that

$$\sum_{m=-M}^{M} e^{imx} = \frac{\sin[x(M+1/2)]}{\sin[x/2]} \to 2\pi\delta(x) \quad (M \to \infty). \tag{12}$$

Note that $\mathbf{k}_1 \neq \mathbf{k}_2$, that is, $\mathbf{k} \neq 0$, since $\mathbf{k}_2$ is in the condensate and $\mathbf{k}_1$ is not in the condensate. Therefore, $\hat{H}_{symm}^{(ext)}$ vanishes for arbitrary BEC in the macroscopically large aperiodic lattice limit whichever order the limits are taken. Therefore, one cannot generate a two-dimensional supersolid via a BEC at temperatures $T \geq 0$ from an external aperiodic lattice potential. However, a two-dimensional supersolid at finite temperatures can be generated via long-range, nonlocal potentials provided by the interparticle interaction which results in self-organization [1], much as Wigner crystallization or Wigner lattice, electrons moving in a uniform background of positive charge that restore electric neutrality [3].

The embedded spaces of D-dimensional quasiperiodic structures are abstract spaces whose dimensions are more than three. The dimensions of the embedded space are dependent on the symmetry of the quasicrystal ($D > 1$) [4, 5]. For example, the quasicrystals with 5, 8-, 10-, and 12-fold symmetry need to be embedded into four-dimensional space, $n = 4$. While for the quasiperiodic structures with 7-, 9-, 18-fold symmetry, the dimension of the embedding spaces increases [4-6] to six, $n = 6$.

The Fibonacci tiling [7, 8] does not fall in the above class of lattice potentials given by (10). However, the Fourier transform of the Fibonacci sequence has δ-function peaks at $k = 2\pi(m + m'\tau)/\sqrt{5}$, where $\tau = (1 + \sqrt{5})/2$ is the golden mean and $m$ and $m'$ are integers [9]. Expressed in terms of Fourier transforms (9) becomes

$$\hat{H}_{symm}^{(ext)} = \frac{\sqrt{N_0}}{V(D)} \sum_{\mathbf{k}'\mathbf{k}} \hat{a}_{\mathbf{k}}^\dagger \xi_{\mathbf{k}'} \tilde{V}_{ext}(\mathbf{k}' - \mathbf{k}) + h.c. \tag{13}$$

where

$$\tilde{V}_{ext}(\mathbf{k}' - \mathbf{k}) = \int d\mathbf{r} V_{ext}(\mathbf{r}) e^{i(\mathbf{k}' - \mathbf{k}) \cdot \mathbf{r}}. \tag{14}$$

Consider the case where $\tilde{V}_{ext}(\mathbf{k}' - \mathbf{k})$ is given by a sum of Dirac δ-functions, which is the case for the Fibonacci tiling [9]. Now the vector $\mathbf{k}' - \mathbf{k}$ must lie either in the condensate or outside the condensate. In either case, $\hat{H}_{symm}^{(ext)}$ vanishes for arbitrary BEC since the vector $\mathbf{k}$ is not in the condensate while the vector $\mathbf{k}'$ is in the condensate.

## 4. Quasicrystal condensate

The necessity that a BEC has the Bloch form and represents a self-organized supersolid for $D \leq 2$ requires that the interaction between the atoms be nonlocal and of infinitely long-range [10]. This proof also applies for the existence of an aperiodic condensate. For instance, macroscopic occupation in the single-particle momenta states $\mathbf{0}$, $\mathbf{q}_1$, $\alpha_1\mathbf{q}_1$, $\mathbf{q}_2$, and $\alpha_2\mathbf{q}_2$, where $\alpha_1$ and $\alpha_2$ are irrational numbers and $\mathbf{q}_1 \times \mathbf{q}_2 \neq \mathbf{0}$, gives rise to additional macroscopic occupation in the single-particle momenta states $(m_1 + \alpha_1 m_2)\mathbf{q}_1 + (n_1 + \alpha_2 n_2)\mathbf{q}_2$, with $m_1, m_2, n_1, n_2 = 0, \pm1, \pm2, \cdots$ owing to the symmetry breaking term $\hat{H}_{symm}$ and the linear momentum conservation of the interparticle potential.

Accordingly, the condensate wave function gets augmented and is given by

$$\psi(\mathbf{r}) = \sqrt{\frac{N_0}{V(D)}} \sum_{m_1,m_2,n_1,n_2=-\infty}^{\infty} \xi_{(m_1+\alpha_1 m_2)\mathbf{q}_1+(n_1+\alpha_2 n_2)\mathbf{q}_2} \, e^{i[(m_1+\alpha_1 m_2)\mathbf{q}_1+(n_1+\alpha_2 n_2)\mathbf{q}_2]\cdot\mathbf{r}}, \qquad (15)$$

where $\mathbf{q}_1$ and $\mathbf{q}_2$ are crystallographic directions.

## 5. Summary and Discussion

We have established that supersolids in $D \leq 2$ cannot be generated via Bose-Einstein condensates in a wide class of quasicrystal potentials that includes the Fibonacci tiling. However, supersolids do arise via Bose-Einstein condensates from infinitely long-range, nonlocal interparticle potentials.